# Dual-Mode Volumetric Optoacoustic and Contrast Enhanced Ultrasound Imaging with Spherical Matrix Arrays

Justine Robin[1,2], Ali Özbek[1,2], Michael Reiss[1,2], Xosé Luis Dean-Ben[1,2], and Daniel Razansky[1,2],*

[1] Institute of Pharmacology and Toxicology and Institute for Biomedical Engineering, Faculty of Medicine, University of Zurich, Switzerland
[2] Institute for Biomedical Engineering, Department of Information Technology and Electrical Engineering, ETH Zurich, Switzerland
* Correspondence to daniel.razansky@uzh.ch

## Abstract

Spherical matrix arrays arguably represent an advantageous tomographic detection geometry for non-invasive deep tissue mapping of vascular networks and oxygenation with volumetric optoacoustic tomography (VOT). Hybridization of VOT with ultrasound (US) imaging remains difficult with this configuration due to the relatively large inter-element pitch of spherical arrays. We suggest a new approach for combining VOT and US contrast-enhanced imaging employing injection of clinically-approved microbubbles. Power Doppler (PD) and US localization imaging were enabled with a sparse US acquisition sequence and model-based inversion based on infimal convolution of total variation (ICTV) regularization. Experiments in tissue-mimicking phantoms and *in vivo* in mice demonstrate the powerful capabilities of the new dual-mode imaging system for blood velocity mapping and anatomical imaging with enhanced resolution and contrast.

## Introduction

The growing use of optoacoustic (OA, photoacoustic) imaging systems in biomedical research has fostered the development of new ultrasound (US) array configurations specifically tailored for an optimal OA imaging performance. Specifically, concave arrays have been shown to provide two- (2D) and three-dimensional (3D) OA images with significantly higher quality than those achieved with standard linear and planar arrays used in pulse-echo US [1], [2]. Volumetric optoacoustic tomography (VOT) with spherical array geometries is generally preferred as it greatly improves accuracy of 3D angiography [3], which has been exploited in both preclinical and clinical studies [4]–[7].
US imaging can provide highly complementary information on tissue morphology and blood flow. In particular, hybrid optoacoustic ultrasound (OPUS) systems based on linear [13]–[16], concave [17] and multi-segment arrays [18] have demonstrated the added value of this multi-modal combination. However, US imaging with the concave spherical arrays commonly employed for VOT applications is hampered by the relatively large element size and inter-element pitch. This leads to high diffraction side lobes preventing proper steering of the US beam, while subsequent US emissions with separate elements results in diminished echo signals, low frame rate, and overall loss of image quality. Microbubbles are FDA-approved contrast agents often employed for enhancing vascular contrast in clinical US imaging applications. Microbubbles have enabled enhancing Power Doppler (PD) images by up to

25 dB [19] and were further used in ultrasound localization microscopy (ULM) as a new super-resolution imaging approach to facilitate accurate blood velocity mapping [20]–[23].

In this work, we propose a new methodology for rendering contrast-enhanced US images with the same spherical matrix array system used to produce the VOT images. PD and localization images from tissue phantoms and mouse brain *in vivo* are shown to provide accurate anatomical reconstructions with as few as 64 detection channels, as well as hemodynamic information complementing those rendered with the VOT modality.

## Materials and Methods

### The imaging system

A custom-made system (Fig. 1a) was used for simultaneous VOT and US imaging. It consists of a spherical array of 512 transducer elements (40 mm radius of curvature, 140° covering angle), with 5 MHz central frequency and >80% detection bandwidth (Imasonic SaS, Voray, France) [24]. It features a central 8 mm aperture for the excitation light delivery through a customized fiber-bundle (CeramOptec GmbH, Bonn, Germany). Each individual transducer element has a diameter of 2.5 mm, separated by 3.2 mm and 3.9 mm pitch form the adjacent elements in the elevational and azimuthal directions, respectively. The array is driven in emission and reception by a custom-made data acquisition system (DAQ, Falkenstein Mikrosysteme GmbH, Taufkirchen, Germany). The DAQ can independently drive all the array elements with digital waveforms of arbitrary duration and arbitrary transmission delays. In receive mode, it can perform parallel digitization of all the detected signals at 12, 24 or 40 $MSps$ and transmit them via Ethernet to a personal computer for storage and post-processing.

A nanosecond-duration optical parametric oscillator (OPO) laser (SpitLight; Innolas Laser GmbH, Krailling, Germany) was used for VOT. It was tuned to 800 nm wavelength and set to 20 Hz pulse repetition frequency (PRF). The DAQ was used to simultaneously process signals acquired from all the 512 transducer elements following every laser pulse (494 samples per channel sampled at 40 $MSps$. The 256 elements located in the central part of the spherical shell were used for US imaging. A two cycle 5 MHz pulse was sequentially emitted by each of these elements in a random order. 254 samples were recorded at 24 $MSps$ sampling rate from randomly selected subsets of transducer elements with increasing sparsity levels. Specifically, 256, 128, 64, 32 or 16 receiving channels were used. The experimental point spread function (PSF) of the system for US imaging was determined by imaging a ~200 $\mu m$ diameter microsphere (BKPMS-1.2 212-250um, Cospheric Inc., Santa Barbara, USA) placed in the centre of the spherical field of view (FOV).

### Micro-vessel phantom

A tissue-mimicking vessel bifurcation phantom was manufactured with agar gel (1.3 % w/v agar powder in water). Two 150 μ$m$ diameter nylon threads were glued together at one extremity, to form a ~400 μ$m$ thread separating into two 150 μ$m$ branches. This structure was fixed into a box where melted agar was poured. After gel solidification, the nylon threads were removed, in a way that channels forming a bifurcation phantom within the agar were produced. A Sonovue microbubble solution (Bracco, Milan, Italy - 1/100 v/v dilution) was flown through the phantom at a rate of 20 μL/min with a syringe pump.

*Animal handling*

Female athymic nude mice (n = 4, 4 to 8 weeks old, Envigo, Netherlands) were used for *in vivo* experiments. Mice were anaesthetized with isoflurane (3% v/v for induction and 1.8 % during the experiments, Provet AG, Lyssach, Switzerland) in an oxygen/air mixture (200/800 mL/min). VOT and US imaging of the brain region was performed with the head of the mouse fixed into a custom-designed stereotactic mouse head holder coupled to a breathing mask (Narishige International, Japan). Blood oxygen saturation, heart rate and body temperature were continuously monitored (PhysioSuite, Kent Scientific, Torrington, USA). The temperature was maintained at ~37°C with a heating pad. A bolus of $40\ \mu L$ of Sonovue microbubble solution (Bracco, Milan, Italy), was injected through the tail vein immediately before each US acquisition session. Two successive acquisitions were made for each mouse, corresponding to a total duration of 120 s. Animals were housed in ventilated cages inside a temperature-controlled room, under a reversed 12-hour dark/light cycle. Pelleted food and water were provided *ad-libitum*. All experiments were performed in accordance with the Swiss Federal Act on Animal Protection and were approved by the Cantonal Veterinary Office Zürich.

*VOT image reconstruction*

VOT image reconstruction was performed with a GPU-accelerated three-dimensional back-projection algorithm [25], [26]. Prior to the reconstruction, raw signals were band-pass-filtered in the 0.1 – 7 MHz range. An ROI covering a $10\ x\ 10\ x\ 4\ mm^3$ volume (100x100x40 voxels) was considered. The spatial resolution achieved with this imaging system was previously reported to lie in the 120-180 µm and 180-280 µm ranges in the axial and lateral directions, respectively [27].

*Power-Doppler US reconstructions*

The procedure to form US images of flowing microbubbles involved several steps. The raw signals were first filtered in the 3-7 MHz range. A singular value decomposition (SVD) clutter filter was then used to isolate the dynamically changing signal of the microbubbles from the static background. The first 10 or 20 eigenvectors out of 100 were removed for phantom and mouse acquisitions, respectively. Two types of algorithms were then used to reconstruct the volumes of interest: a standard delay and sum (DAS) beamformer, and a linear model-based reconstruction algorithm with infimal convolution of two total variation functionals (ICTV) regularization. In both cases, each final frame was obtained by compounding 16 consecutive emissions (among the 256 different transmit events), and a PD image was obtained by summation of all frames in a 30 s acquisition.

The forward model linking the radio frequency (RF) pressure (pulse-echo US) signals measured by the transducer to the distribution of back-scatterers in the imaged medium can be expressed as

$$p(t, r_k) = A[h(r)], \qquad (1)$$

where $p(t, r_k)$ is the signal received on the k-th transducer element at time t, $A$ is a linear operator representing US wave propagation and the impulse response of the transducer, and $h(r)$ represents the back-scattering coefficient of the imaged sample. The inverse problem for US reconstruction is defined from the discretized version of this model, which is written in matrix form as

$$p = Ah, \qquad (2)$$

where $p^{m\times 1}$ is the collection of the time discrete pressure signals from all transducer elements ( m = number of time points × number of transducer elements ), $h^{n\times 1}$ represents the imaged medium discretized on a reconstruction grid (n = number of voxels in the grid), and $A^{m\times n}$ is a model matrix representing the linear operator for this reconstruction problem. The model matrix is associated to the arrangement of sensors, acoustic properties of the medium and US transmission schemes. The i − th column of A contains the signals received on all transducer elements for a point scatterer located in voxel i. Theoretical models [28], simulations [29], or experimental approaches [30], [31] can be used to build this matrix. In this work, the Point Spread Function (PSF) of the imaging sequence was first measured experimentally via DAS beamforming. The PSF for a representative microbubble close to the centre of the array was considered to be the same for the whole field of view (FOV) by applying the corresponding shift. The model-matrix was built with the RF signals corresponding to this PSF accordingly delayed to take into account propagation times from each voxels to the transducer elements. Specifically, a homogeneous speed of sound (c = 1500 m/s) was considered for the whole imaging medium, and the delay laws corresponding to transmission by transducer element i, reception on transducer element j and voxel k were calculated as in the classical DAS beamformer, i.e.,

$$\tau_{ijk} = \frac{d_{ik}+ d_{kj}}{c}, \quad (3)$$

where d represents the distance.

The model inversion was based on ICTV. Total variation regularised model-based reconstruction methods were shown to excel at producing high quality reconstructions from subsampled data sets [32], [33]. Furthermore, ICTV was recently shown to be able to optimally regularise volumetric multi-frame reconstructions accounting for both the spatial and temporal sparsity of the image sequence [34]. The ICTV regularization term was adapted to account for the diverse spatio-temporal information density of the data for the corresponding inversion problem [35], which was defined as

$$h = \underset{h'}{\mathrm{argmin}} \left\{ \frac{1}{2}\|p - Ah'\|_2^2 + \lambda \cdot ICTV(h') \right\}, \quad (4)$$

where $\frac{1}{2}\|p - Ah'\|_2^2$ acts as the fidelity term, $ICTV(h')$ is the regularization term and $\lambda$ is the weighing factor for the regularization term. The $ICTV$ functional is composed of the infimal convolution of two total variation norms $TV_s$ and $TV_t$ weighed differently in spatial and temporal dimensions, i.e.,

$$ICTV(u) = \min_v \{TV_s(u-v) + \gamma \cdot TV_t(v)\} \quad (5)$$

where

$$TV_s(u) = \left\| \sqrt{\left(\frac{\delta u}{\delta x}\right)^2 + \left(\frac{\delta u}{\delta y}\right)^2 + \left(\frac{\delta u}{\delta z}\right)^2 + \kappa^2 \left(\frac{\delta u}{\delta t}\right)^2} \right\|_1$$

and

$$TV_t(h) = \left\| \sqrt{\kappa^2 \left(\frac{\delta u}{\delta x}\right)^2 + \kappa^2 \left(\frac{\delta u}{\delta y}\right)^2 + \kappa^2 \left(\frac{\delta u}{\delta z}\right)^2 + \left(\frac{\delta u}{\delta t}\right)^2} \right\|_1$$

where $\kappa$ is a weighing factor between the two regularisation methods $TV_s$ and $TV_t$ and γ is a parameter allowing for a bias towards either spatial or temporal regularisation. A fast iterative shrinkage and thresholding algorithm (FISTA) [36] was used to split the inversion problem (5) into two parts. The first part involves a gradient descent step for inverting the model matrix, which comprises the dominant processing power requirement of the algorithm. Second part involves computing the proximal operator for the ICTV functional, which is a subproblem solution of an approximation using the primal dual hybrid gradient algorithm [37]. It has been shown that an approximation of the subproblem is enough for convergence [38]. The matrix multiplications involving the model matrix as well as the computation of the proximal operator for ICTV functional were accelerated using a custom OpenCL code running on a GeForce GTX 1060 GPU . The rest of the computations were performed on an AMD Ryzen 5 2600 CPU.

*Microbubble localization and tracking process*

Isolated microbubbles are strong scatterers whose size typically lies well below resolution of the imaging system. Reconstructed images of individual microbubbles thus effectively represent the local PSF of the system. In each reconstructed frame, local intensity maxima were detected, and small regions around these maxima were correlated to such experimentally determined PSF. Note that the same empirical PSF was used over the entire reconstructed volume. Local maxima with correlation coefficients above 0.6 were considered to represent actual microbubbles. Localization of these microbubbles was then further refined using a local quadratic fitting of the intensity maximum, and their positions were stored. A particle tracking algorithm was then used on these positions (simpletracker.m available on mathworks ©Jean-Yves Tinevez, 2019, wrapping matlab munkres algorithm implementation of ©Yi Cao 2009), to track the microbubbles over consecutive frames. Only points corresponding to a maximal linking distance of 0.4 mm (maximum particle velocity of 40 mm/s) were selected.

# Results

The basic system's capability to provide high quality PD images of flow while successfully tracking individual microbubbles was first tested. For this, experiments with sparsely distributed microbubbles flowing in a bifurcation flow phantom were performed to find the proper balance between image quality and frame rate (FR). The FR can either be increased by decreasing the number of transmission events compounded in one frame or the number of receiving transducer elements – this in turn reduces the amount of data collected allowing for a higher PRF. However, quality of images representing individual microbubbles (local system PSF) degrades when the amount of data is reduced (Fig.2). UL imaging is possible when the PSF side lobe intensities decay by 15 dB with respect to the main lobe [39]. The results displayed in Fig. 2 indicate that the maximum FR that can be achieved while fulfilling this criterion is 100 Hz for 16 compounded transmit events and 64 receiving channels.

We subsequently performed experiments in a bifurcation phantom to confirm that such a sparse imaging scheme allows microbubble imaging and tracking with a spherical array transducer. Specifically, we imaged the phantom while flowing a 1/100 v/v diluted Sonovue solution through the channels. To test different levels of sparsity, random subsets of decreasing number of transducer elements (256 to 16 out of 256) were used in receive, and the images were reconstructed using 16 transmit events in coherent compounding with: 1) a classical DAS beamformer, and 2) the model based reconstruction scheme introduced herein (Fig. 3a and 3b, respectively). PD images were then obtained as the mean image averaged over 30 seconds of acquisition. The three channels of the bifurcation are visible in the PD images obtained with both reconstruction methods, although MB – ICTV reconstruction clearly outperforms DAS in terms of CNR (up to 10 dB improvement, Fig. 3, c). In particular, a 5 dB improvement in CNR was achieved for the sparse sequence selected above (16 Tx events and 64 Rx channels). The resolution is also shown to be improved with MB – ICTV. For example, the measured full width at half maximum (FWHM) of the channel was $150\ \mu m$, as expected from the phantom fabrication procedure, even with as few as 16 receiving channels. On the contrary, the equivalent FWHM measured in the DAS reconstructions was $250\ \mu m$ when using 256 to 32 receiving channels, and a $650\ \mu m$ for 16 receiving channels (see Fig. 3 d).

Microbubble localization and tracking was also possible with this sparse sequence. In particular, bubble density and dynamic flow maps were retrieved (Fig. 4). The measured width of the left and right secondary channels was $160\ \mu m$ and $180\ \mu m$, respectively, and the smallest measurable distance between channels was $80\ \mu m$. The respective measured mean velocities in the secondary branches were $7.5 \pm 0.7\ mm/s$ and $5.2 \pm 0.8\ mm/s$, respectively. Considering a small fluid loss at the syringe-phantom connection, this is in good agreement with the $8.2\ mm/s$ and $6.4\ mm/s$ expected velocity values for the measured channel widths. The calculated flow direction map is also in good agreement with the phantom's geometry.

The multimodal imaging capabilities of the system were eventually tested by *in vivo* mouse brain imaging both in the VOT and US modes. PD image quality was first assessed (Fig. 5), and compared to the VOT image for reference. The whole brain was visible through the intact scalp and skull with both modalities. However, a larger number of vessels were distinguished in the US image, especially in the cortex. In particular, $\sim 200\ \mu m$ wide penetrating arterioles could be visualized in the coronal slice (white arrows in Fig. 5a), which are not visible in VOT images arguably due to limited view effects [8]. The smallest measurable distance between vessels was in the order of $200\ \mu m$. Note that in this more complex object the sparse sequence consisting of 16 transmit events and 64 receiving channels only provides a 10 to 12 dB dynamic range in the PD image, hindering proper visualization of vessels below the cortex surface (Fig. 5c). For this reason, we used instead 128 receiving channels for microbubble localization and tracking, resulting in a 50 Hz compounded frame rate. Fig. 6 shows an example of combined VOT, PD image and dynamic flow information obtained through microbubble tracking in a different animal, with a total measurement time below 5 minutes. In the microbubble density and velocity maps, only the bubbles that could be tracked over more than 5 consecutive frames having >0.4 correlation to the PSF are displayed. At 50Hz frame rate microbubble tracking was possible up to a flow velocity of 12 mm/s (Fig. 6d). The measured blood velocities are in agreement with what has been previously reported in the literature [40], [41].

## Discussion and Conclusions

The experimental results obtained in this work demonstrate that VOT imaging systems based on spherical matrix arrays can readily be adapted for dual-mode operation incorporating US contrast-enhanced imaging. Injection of microbubbles facilitated the development of an efficient PD imaging protocol providing accurate vascular features compensating for the limited-view artifacts in VOT images [8]. Localization and tracking of individual microbubbles further enabled enhancing the spatial resolution and rendering flow velocity maps. Sparse US acquisition sequences were adapted for the given distribution of detection elements. Furthermore, phantom experiments demonstrated that the developed model-based (MB – ICTV) reconstruction scheme provides at least a 5 dB improvement over classic DAS beamforming with as few as 16 transmitting and receiving channels. This allowed achieving the image quality and frame rate needed to successfully image and track single microbubbles, while considerably reducing the amount of data being collected with respect to conventional 3D US imaging systems [42].

Microbubbles are routinely-used FDA-approved contrast agents. Thereby, the demonstrated advantages of the new multi-modal imaging approach can be exploited in many biomedical applications and, potentially, also in the clinical setting. Particularly, blood flow mapping and super-resolution imaging can complement the unique functional imaging capabilities of VOT and provide new insights in oncology, cardiovascular biology, metabolic studies and other fields [43]–[46]. The developed methodology is of particular relevance for neuroimaging applications. The ability to obtain diverse structural and functional information at the whole-brain scale with sufficient spatio-temporal resolution is key to understanding the brain function in health and disease. Stand-alone US and VOT systems are independently used for pre-clinical neuroimaging applications yet can provide highly complementary information. In particular, multi-spectral optoacoustic tomography (MSOT) exploits the rich spectroscopic optical contrast in deep tissues to resolve the bio-distribution of oxygenated and reduced hemoglobin, thus providing otherwise unavailable functional and metabolic information from the blood oxygenation readings [44], [47]–[49]. The contrast and resolution however diminish when imaging deeper into the brain, whilst reliable blood flow velocity *in vivo* mapping has not yet been possible with OA modalities [10]–[12]. In contrast, US imaging can measure tissue motion and blood flow dynamics with superior penetration depth of at least several cm. Recently, ULM pushed one step further the hemodynamic readouts that can be provided with US by rendering high quality maps of the brain vasculature and blood velocities with a resolution beating the diffraction limit by an order of magnitude [20]–[23], [50]–[53]. On the other hand, ULM provides limited functional information on brain activity [54], which can be significantly augmented via hybridization with VOT or MSOT.

3D mouse brain vascular imaging through the intact scalp and skull was possible with 200 μm resolution, while blood flow could be mapped for velocities up to 12 mm/s. 3D ULM has previously been achieved by using flat (linear or matrix) array probes with spatial resolution in the $20 - 80\ \mu m$ range [21], [55], [56]. This was achieved however at the cost of either using costly ultrafast parallel acquisition systems featuring 512 to 1024 channels or otherwise opting for long scanning times. The lower resolution of the microbubble density and velocity maps rendered in this work is partially attributed to the relatively low number of individual microbubbles that could be captured for the given limited acquisition durations (120 s), as tens of minutes are usually needed to cover the entire capillary network [57]. Moreover, the localization method could have missed part of the bubbles within the FOV (particularly those close to the image boundaries) due to the fact that the PSF was approximated as constant while local maxima exhibiting a poor correlation with this PSF were discarded. Nevertheless, the spherical matrix array geometry and the developed sparse sequences and reconstruction schemes were shown capable of high resolution imaging with 128 to 256 independent channels, which can be realized with standard (commercial) US electronics [58].

The developed platform further establishes a solid basis for future hardware and software advancements that can greatly mitigate the aforementioned limitations of the current hybrid imaging system implementation. For example, a higher frame rate - achievable with more advanced electronics - would not only allow the measurement of higher blood velocities, but also improve the overall image quality. The MB - ICTV reconstruction scheme was developed to compensate for the motion of microbubbles during the time span required for coherently compounded transmissions [35], thus further improvements in modelling and/or optimization of the regularization terms may contribute to enhancing its performance. The final resolution of the images, in particular the precision of resolving small microvessels and flow profiles, directly depends upon the total number of localized microbubble [57]. It is therefore anticipated that higher frame rates and optimized reconstruction algorithms may significantly boost performance of the suggested approach, both in terms of the anatomical image quality and dynamic flow information.

In conclusion, the new dual-mode VOT and US imaging approach introduced herein efficiently augments the anatomical and functional information provided by stand-alone modalities. The enhanced imaging performance was achieved with a widely employed FDA-approved US contrast agent (microbubbles), which anticipates a large range of applications. Sparse US acquisition sequences and reconstruction schemes were specifically developed around spherical array geometry known to provide optimal VOT imaging performance. The experimental results demonstrate the powerful capabilities of the new dual-mode imaging system for blood velocity mapping and anatomical imaging with enhanced resolution and contrast.

**Acknowledgments**

This project has received funding from the European Research Council (ERC-2015-CoG-682379). The ETH Zurich Postdoctoral Fellowship to J.R. is also acknowledged. The authors would like to thank Baptiste Heiles for fruitful discussions particularly useful for the experimental setup.

FIGURES

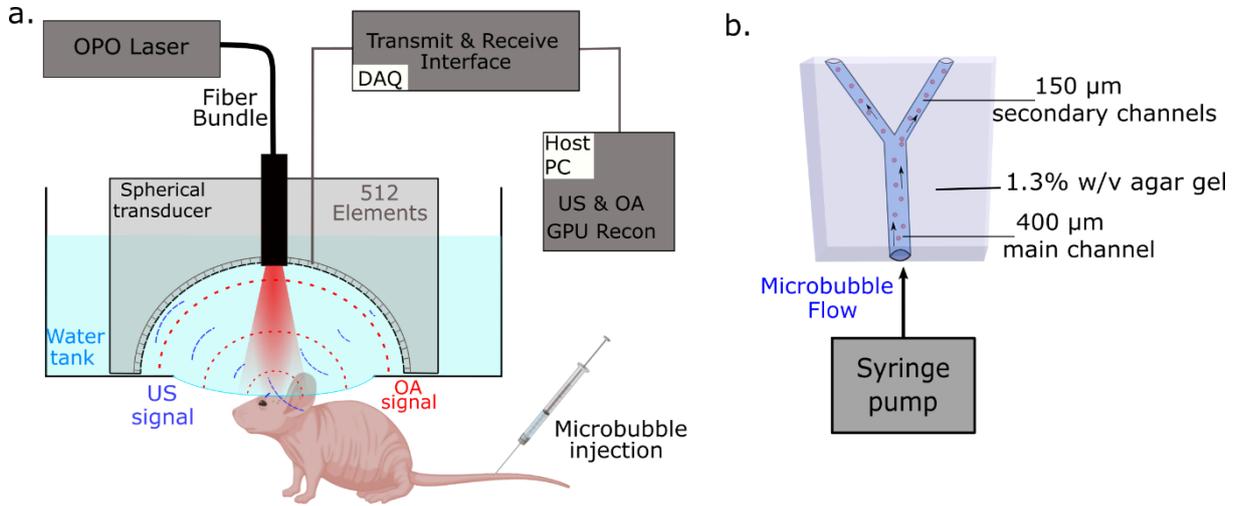

*Figure 1: Experimental Setup*. (a) Sketch of the dual-mode volumetric optoacoustic tomography (VOT) and contrast-enhanced US imaging setup for in vivo imaging of mice. (b) Sketch of the bifurcation flow phantom used for the image quality assessment.

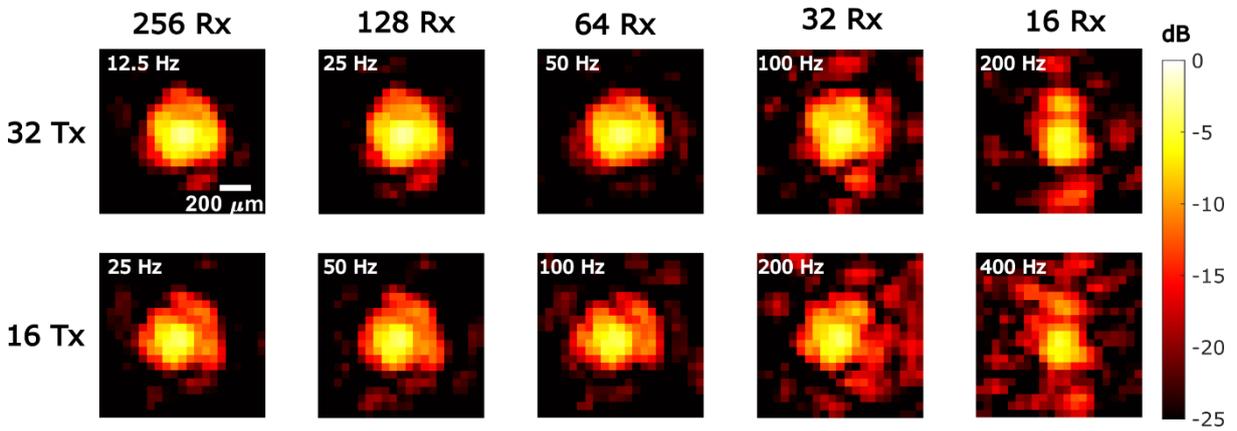

Figure 2: Influence of decreasing number of compounded transmission (Tx) events and receiving (Rx) transducer elements on the experimental PSF and frame rates obtained.

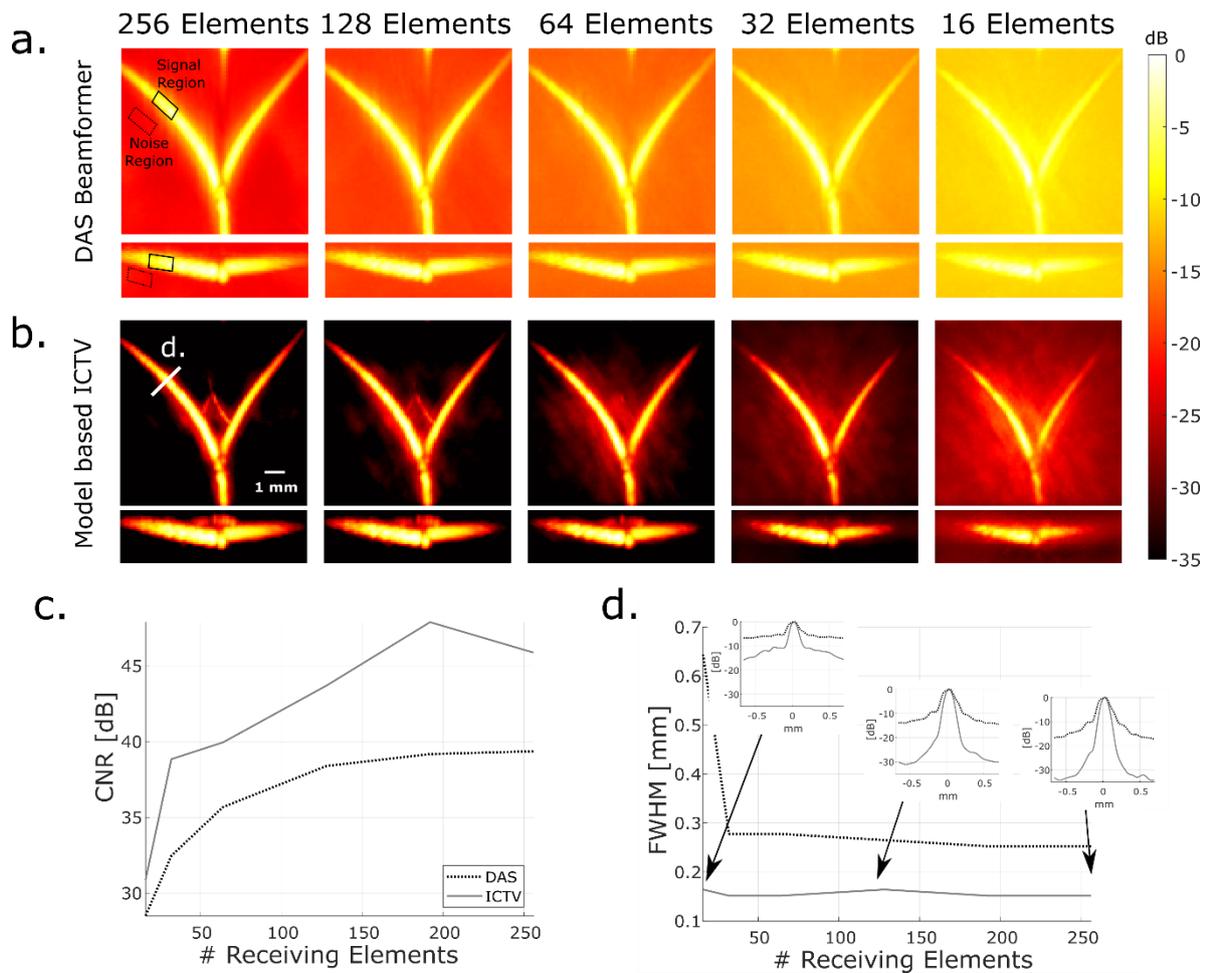

Figure 3: **Sparse imaging of a bifurcation flow phantom**. (a) Power Doppler (PD) image obtained with delay-and-sum (DAS) beamforming for decreasing number of receiving channels – MIPs in the z and y directions are shown. (b) PD image obtained with model based reconstruction with infimal convolution of two total variation functionals (ICTV) for decreasing number of receiving channels – MIPs in the z and y directions are shown. (c) Contrast to noise ratio (CNR) calculated in the phantom's left branch for both reconstruction methods and different numbers of receiving channels. The selected signal and noise regions for CNR calculation are indicated in panel (a). (d) Full Width at Half Maximum (FWHM) of the phantom's left branch for both reconstruction methods and different numbers of receiving channels. Insets show the vessel profile along the segment indicated in panel (b).

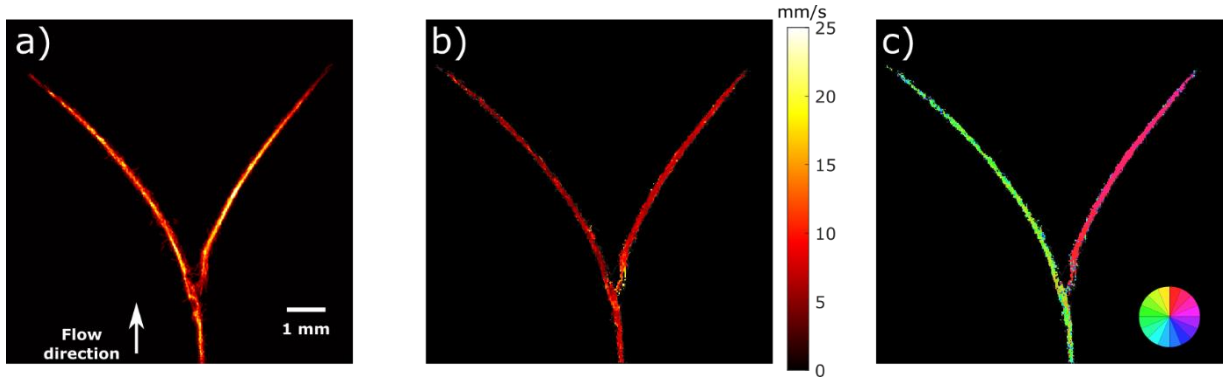

*Figure 4:* **Microbubble US localization images.** *(a) Microbubble density map, MIP in the z direction is shown. (b) Flow velocity map, MIP in the z direction is shown. (c) Representation of the flow direction in the phantom. Each colour corresponds to one specific flow direction, as indicated by the colour wheel. Only 2D flow in the X-Y plane is shown.*

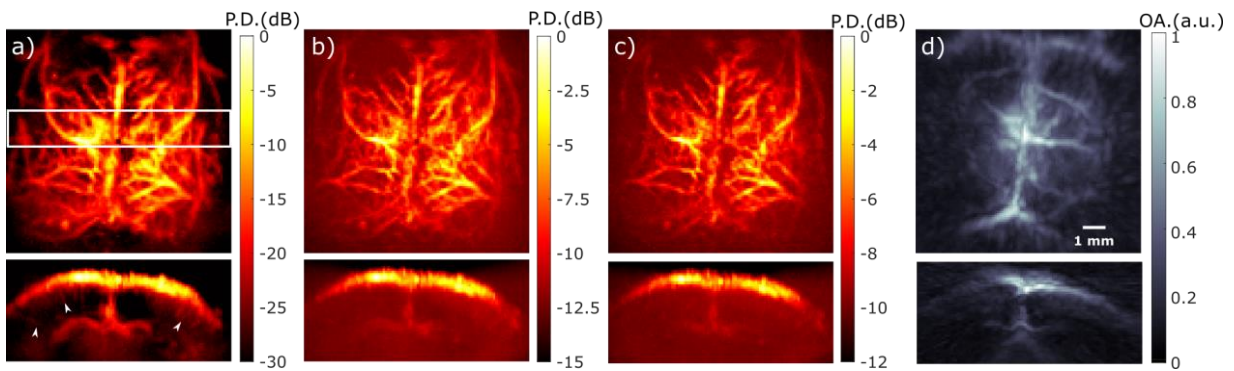

*Figure 5:* **Power Doppler imaging of a whole mouse brain with a spherical matrix array transducer through intact scalp and skull.** *(a)* *PD image, obtained in a 1 min acquisition sequence with 16 Tx events and 256 receiving channels. Top: axial view, MIP in the z direction, bottom: coronal view, 2 mm slice corresponding to the white box displayed in panel (a).* *(b)* *PD image, obtained in a 1 min acquisition sequence with 16 Tx events and 128 receiving channels. Top: axial view, MIP in the z direction, bottom: coronal view, 2 mm slice corresponding to the white box displayed in panel (a).* *(c)* *PD image, obtained in a 1 min acquisition sequence with 16 Tx events and 64 receiving channels. Top: axial view, MIP in the z direction, bottom: coronal view, 2 mm slice corresponding to the white box displayed in panel (a).* *(d)* *Reference VOT image at 800 nm. Top: axial view, MIP in the z direction, bottom: coronal view, 2 mm slice corresponding to the white box displayed in panel (a).*

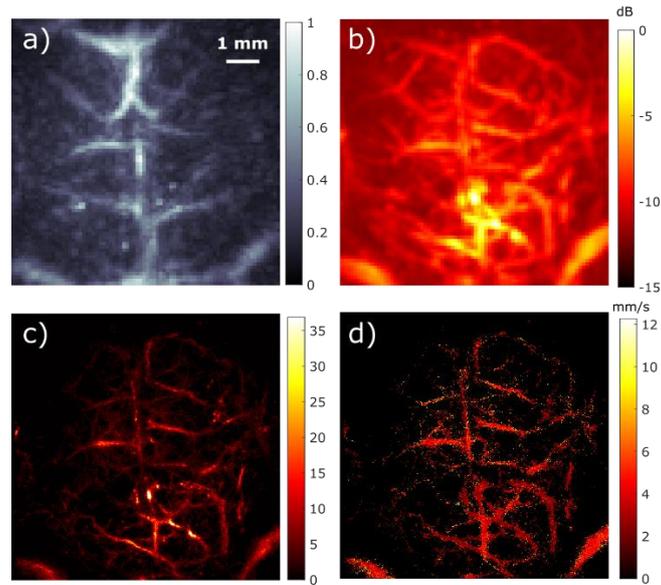

*Figure 6:* **Multi-modal VOT and US imaging of the mouse brain**. *(a) VOT image of a whole mouse brain at 800 nm. MIP along the z direction is shown. (b) PD image obtained in a 1 min acquisition sequence with 16 Tx events and 128 receiving channels. (c) Microbubble density map obtained in a 2 min acquisition sequence. Only bubbles exhibiting a correlation with the experimental PSF > 0.4 and trajectories longer than 5 frames are displayed. (d) Bubble velocity map. Median along the z direction is shown. Only bubbles exhibiting a correlation with the experimental PSF > 0.4 and trajectories longer than 5 frames are displayed.*